\documentclass[aps,11pt]{revtex4}

\begin{document}

\author{Yakir Aharonov\thanks{Electronic address: yakir@post.tau.ac.il}}
\address{School of Physics and Astronomy, Tel-Aviv University,
Israel \\ and \\ Chapman University, Orange, CA} 
\author{Pawel O. Mazur\thanks{Electronic address: mazur@physics.sc.edu} }
\address{Department of Physics and Astronomy, 
\\ University of South Carolina, Columbia, SC 29208}
\title{Stability of Exponential Tails in the Scattering on Wedges and Impurities}
\date{April 13, 1999}

\begin{abstract}
We study a simple exactly solvable 2D model describing the interaction of a
localized particle with an impurity. The localization potential $V(x)=-\alpha \delta (x)$ 
causes the particle to be trapped in the $y$-axis,
and the `impurity' is modeled by a straight impenetrable edge extending along
the positive $x$-axis. We show that the problem described can be treated as
the Sommerfeld diffraction from an infinite edge. We use the model to
present qualitative arguments on the nature of interaction of polarized
light travelling along an optical fiber with external impurities.
\end{abstract}
\maketitle

\section{Introduction}

There is a class of problems characterized by an example of polarized light
propagating along an optical fiber (or a wave guide). It is interesting to
see how the interaction with depolarizing impurities arises in such a
problem and what effect the impurities have on the propagation of polarized
light. The polarized light is described by the scalar Helmholtz equation.
The optical fiber is a waveguide with appropriate boundary conditions
imposed. The depolarizing interaction with impurities on the walls of the
waveguide can be modeled by a photon spin-flip interaction. Since this
problem is too complex to obtain an exact analytic solution, we will
consider a much simplified model which, we believe, captures the essense of
the problem. The Schr\"{o}dinger equation for a particle of unit mass $m=1/2$
is the same as the Helmholtz equation. We will model the waveguide by an
attractive, localizing, potential $V(x,y)$ and the impurity by an
impenetrable barrier which we will take to be a semi-infinite edge located
along the positive $x$-axis\footnote{
More generally, we will assume that it extends from $x=a\geq 0$ to infinity.}.
A simple argument based on the first order perturbation theory (the Born
approximation) applied to a bound state $\psi _{0}$ in a potential $V(x,y)$
suggests that an ``impurity'' located on the $x$-axis at $x=a$, and modeled
by a potential $V_{imp}(x,0)=\lambda \delta (x-a)\delta (y)$, modifies the
bound state wave function by a correction which is quadratic in the value of
this wave function at the point where the ``impurity'' is located. What this
means is that in the case when the bound state wave function decays
exponentially with $x$ as $|\psi _{0}(x,y)|$ $\sim e^{-\alpha |x|}$, the
correction to the wave function caused by an ``impurity'' is of the
subleading order, $|\psi ^{(1)}(x,y)|$ $\sim $ $e^{-2\alpha a}e^{-\alpha
|x|} $. In this sense the bound state `exponential tails' are rather rigid
because the ``impurity'' changes the asymptotic behavior of the bound state
by an exponentially negligible correction to an overall normalization factor.

\section{\textrm{The Model }}

Consider a particle moving in the Euclidean plane. Along the positive $x$%
-axis we place an impenetrable barrier, a straight semi-infinite edge. A
free quantum mechanical particle will be scattered by a straight edge in a
manner first described by Sommerfeld \cite{Somm1,Somm2,Pauli,Morette}. We
are analyzing a situation where the particle is trapped in an attractive
potential $V(x,y)=-\alpha \delta (x)$ that allows it to move freely only in
the $y$ direction. The Hamiltonian for this particle is the sum of two
one-dimensional Hamiltonians $H=H_{1}+H_{2}$ , $H_{1}=-\partial
_{x}^{2}-\alpha \delta (x)$, $H_{2}=-\partial _{y}^{2}$ , where $\alpha >0$.
Therefore, the total Hamiltonian is

\begin{equation}
H=-\left( \partial _{x}^{2}+\partial _{y}^{2}\right) -\alpha \delta
(x).\smallskip  \label{hamil}
\end{equation}
In the simplest possible case that we deal with here, $H_{1}$ has only one
bound state,

\begin{equation}
\psi _{0}(x)=e^{-\alpha |x|}\text{ .}  \label{bound1}
\end{equation}
In addition, there are scattering states,

\begin{equation}
\psi (x)=e^{ipx}+Ae^{-ipx},x>0,\psi (x)=Be^{ipx},x<0,  \label{scatt}
\end{equation}
where the transmission and reflection coefficients are

\begin{equation}
A=\frac{i\alpha }{p-i\alpha }\text{ , }B=\frac{p}{p-i\alpha }\text{ .}
\label{transmit}
\end{equation}
A simple pole at $p=i\alpha $ in the S-matrix corresponds to a bound state (%
\ref{bound1}).

The stationary solution of the Schr\"{o}dinger equation describing a bound
state of $H$ and corresponding to the energy

\begin{equation}
E=k^{2}-\alpha ^{2},  \label{energy}
\end{equation}
is

\begin{equation}
\psi _{0}(x,y)=e^{iky-\alpha \left| x\right| }.  \label{bound2}
\end{equation}

The problem we want to solve is the problem of scattering on the straight
edge $x>a$, $y=0.$ We assume that for $x>a$, $y=0$ the wave function or its
normal derivative must vanish, thus either $\psi (x,y)=0$ or $\partial
_{y}\psi (x,y)=0.$ These conditions are equivalent to the Dirichlet or
Neumann boundary conditions, correspondingly. The idea that underlies the
solution is to explore the Helmholtz equation
\begin{equation}
\left( \Delta +k^{2}\right) \psi =0,  \label{helmho}
\end{equation}
in the complex $x$-$y$ space (complex $x$ and $y$ planes), where
\begin{equation}
\Delta =\partial _{x}^{2}+\partial _{y}^{2},  \label{lapla1}
\end{equation}
and $x$ and $y$ are independent complex variables. Elementary solutions that
represent plane waves $\exp (ik_{1}x+ik_{2}y)$ are holomorphic everywhere in
the $x$ and $y$ complex planes. Since the Helmholtz equation is linear, the superposition
principle can be used to obtain more general solutions from the basic ones.
In particular, the superposition principle can be applied to obtain bound
state solutions.

Now,

\begin{equation}
\psi _{0}=A\exp (-ikr\cos (\varphi -\beta _{0})),
\end{equation}
\begin{equation}
\psi =\int_{C}A(\beta )\exp (-ikr\cos (\varphi -\beta ))d\beta .
\end{equation}
To get $\psi _{0}$ from the last formula we need to choose:

\begin{equation}
A(\beta )=\exp (i\beta )/2\pi (\exp (i\beta )-\exp (i\beta _{0}))
\end{equation}
and so
\begin{equation}
\psi =\frac{1}{4\pi }\int_{C}\frac{\exp (i\gamma /2)\exp (ikr\cos \gamma )}{%
(\exp (i\gamma /2)-\exp (-i\chi /2))}d\gamma ,
\end{equation}
where $\chi =(\varphi -\beta _{0})/2$. For the scattering states arising in
the scattering off the straight edge, $\beta _{0}=\pi /2$. For the bound
state propagating along the $y$-axis only, we have $\beta _{0}=\pi /2\pm
i\lambda $, where $\lambda =\lambda (k,\alpha )$.

The cut starting at $x=a$ and running all the way in the positive direction
of $x$-axis suggests introducing a double covering of a complex $z$-plane,
where $z=y+i(x-a)$. To this end, we introduce a new complex variable $w=\xi
+i\eta $ defined by
\begin{equation}
z=w^{2}.  \label{cover}
\end{equation}
We can now write
\begin{equation}
x-a=r\cos \varphi =-r\sin (\varphi -\pi /2),
\end{equation}
\begin{equation}
y=r\sin \varphi =r\cos (\varphi -\pi /2),
\end{equation}
and identify $\xi $ and $\eta $ as
\begin{equation}
\xi =\sqrt{r}\cos \frac{\chi }{2},\eta =-\sqrt{r}\sin \frac{\chi }{2},
\label{newangle}
\end{equation}
where $\chi =\varphi -\pi /2$. We observe that both $\varphi =0$ and $2\pi $
correspond to $y=0$ and $x>a$ (as $x=r+a$), but the first case translates
into $\xi =\eta =\sqrt{r/2}$, whereas the second one into $\xi =\eta =-\sqrt{%
r/2}$. For $\varphi =\pi $ we obtain that $\xi =-\eta =\sqrt{r/2}$. This
corresponds to $y=0$ but with $x<a$. In these new variables the Laplacian
reads
\begin{equation}
\Delta =\frac{1}{4(\xi ^{2}+\eta ^{2})}(\partial _{\xi }^{2}+\partial _{\eta
}^{2}).  \label{lapla2}
\end{equation}
We will seek the solution to the scattering problem described by the
Helmholtz equation (\ref{helmho}) in the form of $\psi =\psi _{1}+\psi _{2}$
, where
\begin{equation}
\psi _{1}=\exp (-iky)V(\xi ,\eta ),
\end{equation}
\begin{equation}
\psi _{2}=\exp (iky)U(\xi ,\eta ).
\end{equation}
Plugging $\psi _{1}$ into (\ref{helmho}) yields
\begin{equation}
\left[ (\partial _{\xi }^{2}+\partial _{\eta }^{2})-4ik(\xi \partial _{\xi
}-\eta \partial _{\eta })\right] V(\xi ,\eta )=0,
\end{equation}
and suggests that one should take $V(\xi ,\eta )=V(\xi )$ and $U(\xi ,\eta
)=U(\eta )$. As a result, one obtains
\begin{equation}
V(\xi )=\int_{-\infty }^{\xi }d\tau \exp (2ik\tau ^{2})=F(\xi )=C(\xi
)+iS(\xi ),
\end{equation}
where $C(\xi )$ and $S(\xi )$ are Fresnel integrals. The expression for $%
U(\eta )\,$is identical to that of $V(\xi )$ except that it is in terms of $%
\eta $. Consequently,
\begin{equation}
\psi (\xi ,\eta )=\exp (-iky)F(\xi )-\exp (iky)F(\eta ),
\end{equation}
where, as seen from (\ref{cover}),
\begin{equation}
y=\xi ^{2}-\eta ^{2},\text{ }x=2\xi \eta +a.  \label{relat}
\end{equation}
For $y=0$, we have
\begin{equation}
\psi (x,0)=F(\xi )-F(\eta )
\end{equation}
which equals $0$ for $x>a$ and $\left( F(\xi )-F(-\xi )\right) $ otherwise,
as should be apparent from our earlier discussion of the relationship
between $\xi $ and $\eta $ in the case $y=0$.

To conclude this part, we have demonstrated that the solution to the pure
scattering problem under study in the entire $x$-$y$ plane is given by
\begin{equation}
\psi (x,y)=C_{0}\left[ \exp (-iky)\int_{-\infty }^{\xi }d\tau \exp (2ik\tau
^{2})-\exp (iky)\int_{-\infty }^{\eta }d\tau \exp (2ik\tau ^{2})\right] ,
\end{equation}
where $\xi $ and $\eta $ are related to $x$ and $y$ via (\ref{relat}) and $%
C_{0}$ is a constant.

Let us now consider the case of the bound state and assume for simplicity
that $a=0$. Writing $\psi _{0}(x,y)$ given by (\ref{bound2}) as $\psi
_{0}(x,y)=\exp (iS)$, we obtain
\begin{eqnarray}
iS &=&-iky-\alpha |x|=ikr\sin \varphi -\alpha \varepsilon r\cos \varphi
\nonumber \\
&=&r\left( (k+\alpha \varepsilon )\exp (i\varphi )-(k-\alpha \varepsilon
)\exp (-i\varphi )\right) /2  \nonumber \\
&=&i\kappa r\cos (\varphi -\pi /2-i\lambda )=i\kappa r\cos (\varphi -\chi
_{0})=i\kappa r\cos \chi ,
\end{eqnarray}
where we used the substitutions $k+\alpha \varepsilon =\kappa \exp \lambda $
and $k-\alpha \varepsilon =\kappa \exp (-\lambda )$. Moreover, $\varepsilon
=\varepsilon (x)=x/|x|$ is the signum function and $\chi _{0}=\pi
/2+i\lambda $. The last sequence of formulas is valid for the positive
values of energy. The expression for $iS$ that corresponds to negative
energies can be easily obtained by analytical continuation. By analogy to
the free scattering case, one can introduce new complex variables $\xi $ and
$\eta $ related to the radial and angular coordinates $r$ and $\chi $ in the
same manner as in (\ref{newangle}) except that now
\begin{equation}
\xi =\sqrt{\frac{r}{2}}\left( \cos \frac{\varphi -i\lambda }{2}+\sin \frac{%
\varphi -i\lambda }{2}\right) ,
\end{equation}
\begin{equation}
\eta =\sqrt{\frac{r}{2}}\left( \cos \frac{\varphi -i\lambda }{2}-\sin \frac{%
\varphi -i\lambda }{2}\right) .
\end{equation}
One can show that for $\varphi $ equal $0$ and $2\pi $, $\xi (\varphi )=\eta
^{*}(\varphi )$, where the asterisk denotes the complex conjugation.
Pursuing further the analogy to the free scattering, it is straightforward
to write the solution for the case under consideration,
\begin{equation}
\psi =\exp (-\alpha |x|)\left( \exp (-iky)F(\xi )-\exp (iky)F(\eta
^{*})\right) .
\end{equation}
This is an exact solution to the scattering problem on the half-infinite
edge impurity which is valid both for real and imaginary $\kappa .$ The case
of imaginary $\kappa$,  
or negative energy, is of
interest because it corresponds to the case of a single bound state of the
attrractive Dirac-delta potential in the transverse direction $x$. In this
case we find an agreement with the simple perturbation theory argument
presented in the introduction. Our  investigation can be extended in many
directions. In particular, it seems possible to extend the method presented
here to localized wave packets. \bigskip \bigskip \bigskip \bigskip

\textbf{Aknowledgements\medskip \medskip }

This research was partially supported by the NSF grant to the University of South Carolina and Chapman University.


\begin{thebibliography}{9}
\bibitem{Somm1}  A. Sommerfeld, \emph{``Matematische Theorie der
Diffraction'',} Math. Ann. \textbf{47}, 317-374,\ 1896.

\bibitem{Somm2}  A. Sommerfeld, \emph{Optics, }Academic Press, New York,
1954, pp.247-272.

\bibitem{Pauli}  Pauli, W., \emph{``On Asymptotic Series for Functions in
the Theory of Diffraction of Light''}, Phys. Rev. \textbf{54}, 924-931, 1938.

\bibitem{Morette}  De-Witt-Morette, C., Low, S. G., Schulman, L. S. and
Shiekh, A. Y., `\emph{`Wedges I''}, Foundations of Physics \textbf{16},
311-349, 1986.
\end{thebibliography}
\end{document}